\documentclass[twocolumn,letter]{jpsj3}

\usepackage{graphicx}
\usepackage{dcolumn}
\usepackage{bm}
\usepackage{color}
\usepackage{txfonts}
\def\vector#1{{\boldsymbol{#1}}}
\def\va{{\vector a}}
\def\vb{{\vector b}}
\def\vc{{\vector c}}

\def\vH{{\vector H}}
\def\vk{{\vector k}}

\def\vq{{\vector q}}

\def\Tc{{T_{\rm c}}}

\def\dps{\displaystyle}

\def\TMTSFX{\mbox{${\rm (TMTSF)_2}X$}}
\def\TMTSFClO{\mbox{${\rm (TMTSF)_2ClO_4}$}}

\def\hsp#1{\hspace{#1ex}}

\def\eq.#1{Eq.~(\ref{#1})}
\def\eqs.#1{Eqs.~(\ref{#1})}

\def\Hc2{{H_{\rm c2}}}
\def\difHc2{{H'_{\rm c2}}}

\def\Jerome{J\'{e}rome}

\def\Pevelen{P\'{e}velen}

\def\*ref*{{\color{red}$\leftarrow$Ref.{\bf [~~~~~]}}}

\def\PA{${\rm P}_{\rm A}$}
\def\PB{${\rm P}_{\rm B}$}

\def\Pave{${\rm P}_{\rm ave}$}
\def\Aave{${\rm A}_{\rm ave}$} 
\def\Modified#1{${\rm M}_{\rm #1}$}

\newcommand\Equation[2]{
\begin{equation}\label{#1} 
#2
\end{equation}
}

\title{
Stabilization of the Fulde--Ferrell--Larkin--Ovchinnikov State \\ 
by Tuning In-plane Magnetic-Field Direction: \\ 
Application to a Quasi-One-Dimensional Organic Superconductor
}

\author{Katsumi Itahashi and Hiroshi Shimahara}

\inst{
Department of Quantum Matter Science, ADSM, Hiroshima University, \\ 
Higashi-Hiroshima, Hiroshima 739-8530, Japan
}

\recdate{May 25, 2018}

\abst{
The Fulde--Ferrell--Larkin--Ovchinnikov (FFLO) state 
in quasi-one-dimensional systems with warped Fermi surfaces 
is examined in strong parallel magnetic fields. 
It is shown that 
the state is extremely stable for field directions 
around nontrivial optimum directions, 
at which the upper critical field exhibits cusps, 
and that the stabilization is due to a Fermi-surface effect 
analogous to the nesting effect for the spin density wave 
and charge density wave. 
Interestingly, the behavior with cusps 
is analogous to that in a square lattice system 
in which the hole density is controlled. 
For the organic superconductor {\TMTSFClO}, 
when the hopping parameters obtained by previous authors 
based on X-ray crystallography results 
are assumed, 
the optimum directions are 
in quadrants consistent with the previous experimental observations. 
Furthermore, near this set of parameters, 
we also find sets of hopping parameters 
that more precisely reproduce the observed optimum in-plane field directions. 
These results are consistent with the hypothesis that 
the FFLO state is realized in the organic superconductor. 
}

\begin{document}
\sloppy
\maketitle

The existence of the 
Fulde--Ferrell--Larkin--Ovchinnikov (FFLO) state~\cite{Ful64,Lar64,Cas04} 
in exotic superconductors 
such as heavy fermion~\cite{Mat07} and 
organic~\cite{Leb08} superconductors 
has been suggested. 
In the quasi-one-dimensional (Q1D) 
organic compound \TMTSFClO~\cite{NoteTMTSF}, 
Yonezawa et al. observed the $\phi$ dependence of 
the onset transition temperature 
$T_{\rm c}^{\rm onset}(\phi)$,~\cite{Yon08a,Yon08b} 
where $\phi$ denotes the angle between 
the in-plane magnetic field $\vH$ 
and the crystal a-axis. 
They found that 
the principal axis of $T_{\rm c}^{\rm onset}(\phi)$ changed 
at a high field $H_0$, 
and argued that 
this may be related to the emergence of the FFLO state. 
This motivated us to examine the FFLO state in Q1D systems, 
particularly the dependence of its stability 
on the in-plane magnetic-field direction.

{\it Possible origin of in-plane anisotropy} -- 
The observed change of the principal axis, 
if caused by a transition to the FFLO state, 
must primarily originate from 
the emergence of the nonzero center-of-mass momentum $\vq$ 
of the Cooper pairs, 
which is characteristic of the FFLO state. 
Unless the orbital effect is extremely weak, 
the modulation of the order parameter due to the FFLO state 
(the FFLO modulation) 
can occur in the direction 
parallel to the vortex line~\cite{Gru66,NoteHLL}. 
In {\TMTSFClO}, 
the temperature dependence of 
the upper critical field~\cite{Lee95,Yon12,NoteOrb} 
shows that 
the orbital effect is not negligibly small. 
Therefore, 
it is reasonable to assume that $\vq \parallel \vH$.

Even when $\vq = 0$, 
the transition temperature $T_{\rm c}(\phi)$ 
and upper critical field $H_{\rm c}(\phi)$ 
depend on $\phi$ 
because of the orbital pair-breaking effect, 
and they reflect the anisotropy of the Fermi surface. 
In {\TMTSFX}, 
one of the hopping integrals is much larger than the others, 
which gives rise to a highly conductive chain. 
Therefore, 
the orbital pair-breaking effect must be predominantly determined 
by the magnitude of the component of $\vH$ perpendicular to the chain, 
which is consistent with 
the observed behavior of $T_{\rm c}^{\rm onset}(\phi)$ below $H_0$. 
Therefore, below $H_0$, the anisotropy of 
$T_{\rm c}^{\rm onset}(\phi)$ primarily originates from 
the orbital pair-breaking effect, 
and 
the paramagnetic pair-breaking effect does not significantly 
contribute to the anisotropy. 
In contrast, in the FFLO state, the finite $\vq$ 
gives rise to an additional effect from the Fermi-surface anisotropy 
on the anisotropies of $T_{\rm c}(\phi)$ and $H_{\rm c}(\phi)$. 
Because $\vq \parallel \vH$, 
the dependence on the direction of $\vH$ 
is equivalent to that on the direction of $\vq$. 
When $\vq \ne 0$, 
$\phi_{\vq} = \phi$, 
where $\phi_{\vq}$ denotes the angle between 
$\vq$ and the crystal a-axis.

{\it Nesting effect for FFLO state} -- 
Interestingly, 
most of the candidate compounds for the FFLO state 
are quasi-low-dimensional, 
presumably 
because of 
(i) the suppression of the orbital pair-breaking effect 
in parallel magnetic fields~\cite{Leb08,Shi97b,Shi99b,Shi09,Shi94} 
and 
(ii) a stabilization effect that originates from 
the Fermi-surface structure.~\cite{Shi94,Shi97a,Shi99aT0,Shi99aFT}

Effect (ii) is called the nesting effect for 
the FFLO state~\cite{Shi94} 
in analogy to that for 
the spin and charge density waves (SDW and CDW). 
If two electrons with 
$\vk \uparrow$ and $-\vk + \vq \downarrow$ are simultaneously 
near the Fermi surfaces split by the Zeeman energy, 
they easily form a Cooper pair. 
Therefore, 
if such momenta $\vk$ occupy a large portion of the momentum space, 
the FFLO state is stable. 
The ``measure'' of the occupied portion 
depends on the shapes of the Fermi surfaces 
split by the Zeeman energy, 
and 
is closely related to the stability of the FFLO state. 
In addition, 
the momentum dependence of the gap function near 
the Fermi surfaces must be taken into account.

To examine the nesting effect, 
it is useful to consider 
the overlap of 
the Fermi surface of the up-spin electrons and 
the Fermi surface of the down-spin electrons 
that is shifted by $\vq$ 
(hereafter simply expressed as ``the Fermi surfaces'' 
at some subsequent instances below). 
It is easily found that 
in one-dimensional (1D) systems, 
the Fermi surfaces fully touch, 
where one of them is shifted by an appropriate $\vq$; 
that is, the nesting is perfect, 
which causes 
the upper critical field of the FFLO state 
to diverge at $T = 0$. 
Hence, it appears that the FFLO state is most stable 
in nearly 1D systems.~\cite{NoteN1D} 
However, 
in the nearly 1D system, 
the usual nesting instability induces the SDW or CDW 
at a higher transition temperature 
for realistic coupling constants. 
Therefore, 
quasi-two-dimensional (Q2D) systems 
in which the SDW and CDW transitions are suppressed 
must be most favorable to the FFLO state. 
Note that 
in this context, 
the {\TMTSFX} compounds are classified as Q2D systems 
in the sense that 
the Fermi surfaces are sufficiently warped that 
the SDW instability is suppressed, 
although they are traditionally called ``Q1D'' organic superconductors 
because the hopping integrals in the crystal a-direction are 
much larger than in the other directions.

In this letter, 
we examine the scenario in which 
$T_{\rm c}^{\rm onset}(\phi)$ is maximized 
when $\vq$ is oriented by $\vH$ 
in the optimum direction due to the nesting effect. 
To focus on this effect, 
we ignore the orbital pair-breaking effect. 
Therefore, 
in our theoretical model, 
we assume that the orbital effect 
is sufficiently strong to lock the direction of $\vq$ along $\vH$, 
and 
is negligibly weak 
in the equations for $T_{\rm c}$ and $H_{\rm c}$. 
The latter part of this assumption 
is not quantitatively justified for {\TMTSFClO}; 
however, 
even in our simplified model, 
it would be possible to clarify 
the directions of the magnetic fields 
that most stabilize the FFLO state.

{\it Sensitivity of nesting effect} -- 
Because the nesting effect sensitively depends on 
the Fermi-surface structure, 
a reliable result---even an approximation---for the angular dependence 
cannot be obtained 
solely by simple considerations regarding the shape of the Fermi surface.

To clarify the nesting effect for the FFLO state in detail, 
one of the authors studied 
a superconductor on a square lattice 
using its ability to realize various shapes for the Fermi surface 
by changing the hole density $n_{\rm h}$~\cite{Shi99aT0,Shi99aFT}. 
It might be expected from the analogy with the 1D system 
that the FFLO state is most stable 
when the Fermi surface has flat portions. 
However, in reality, 
a round Fermi surface 
at $n_{\rm h} \approx 0.630$ 
provides the greatest stability for the FFLO state~\cite{Shi99aT0}. 
At this hole density, 
$H_{\rm c}(n_{\rm h})$ exhibits a sharp cusp 
and exceeds five times the Pauli paramagnetic limit.

The sharp cusp in $H_{\rm c}(n_{\rm h})$ can be explained 
as follows.~\cite{Shi99aT0} 
The difference between the Fermi surfaces mentioned above 
can be expressed by 
$\Delta k_{{\rm F}x}(k_y,\vq) 
 \equiv 
   k_{{\rm F}x}^{\downarrow}(k_y - q_y) 
 - k_{{\rm F}x}^{\uparrow}(k_y) + q_x$, 
where 
$k_x = k_{{\rm F}x}^{\sigma}(k_y)$ 
expresses the Fermi surface of $\sigma$-spin electrons. 
We define $k_y^{0}(\vq)$ 
by 
$\Delta k_{{\rm F}x}(k_y^{0}(\vq),\vq) = 0$, 
and suppose that 
$\Delta k_{{\rm F}x}(k_y,\vq)$ is expanded 
as 
$\Delta k_{{\rm F}x}(k_y,\vq) \propto (k_y - k_y^{0})^n$ 
near $k_y = k_y^{0}(\vq)$ 
with an integer $n$. 
The critical field is enhanced for the $\vq$ that gives $n = 2$, 
which implies that the Fermi surfaces touch 
on the line at $k_y = k_y^0(\vq)$. 
In a square lattice system 
at $n_{\rm h} \approx 0.630$, 
$n = 4$ for an appropriate $\vq$, 
which results in the previously mentioned sharp cusp 
and extreme enhancement of $H_{\rm c}(n_{\rm h})$.

{\it Previous theories} -- 
The FFLO state has been theoretically examined 
in {\TMTSFClO} by many authors.~\cite{NoteFFLOTMTSF,
Shi00mix,Leb11,Leb10,Cro12,Cro17,Miy14,Miy16,Aiz09} 
For example, 
Lebed and Wu 
compared their theoretical curve of $H_{\rm c2}(T)$ 
with the experimental data~\cite{Yon08a}, 
and obtained a good overall qualitative and 
quantitative agreement.~\cite{Leb10} 
Croitoru, Houzet, and Buzdin 
studied the interplay between 
the orbital effect and the FFLO modulation, 
and obtained results suggesting that the modulated phase stabilization 
was the origin of the magnetic-field angle dependence of 
$T_{\rm c}^{\rm onset}(\phi)$.~\cite{Cro12} 
Miyawaki and Shimahara 
examined the effect of the Fermi-surface anisotropy 
in Q1D systems,~\cite{Miy14} 
and found a temperature-induced dimensional crossover of $H_{\rm c}(T)$ 
from one dimension to two dimensions, 
which may be related to a small shoulder 
observed in the upper critical field curve 
for $\vH \parallel \va$ in {\TMTSFClO}~\cite{Yon08a,Yon08b}. 
However, 
the relation between the change of the principal axis and 
the Fermi-surface effect mentioned above 
has not been clarified.

{\it Assumptions and model} -- 
In {\TMTSFClO} at ambient pressure, 
the anion order doubles the periodicity in the crystal b-direction, 
and influences 
the electron energy dispersion near the edges 
of the Brillouin zone halved by the anion order. 
This also corrects the quasi-particle excitations in superconductors 
and may eliminate 
the line nodes of the d-wave gap function~\cite{Shi00ao}; 
however, 
it does not significantly affect the FFLO state, 
because, as shown below, 
the optimum $\vq$ makes the Fermi surfaces touch 
at $k_y$ that is far away from those edges~\cite{NoteMiy16}. 
Therefore, we neglect the effect 
of the anion order on the FFLO state.

Information on the crystal structure is indispensable 
for a close comparison between the theoretical and experimental results. 
We adopt the cell parameters 
at a low temperature and under atmospheric pressure 
obtained by {\Pevelen} et al.~\cite{Pev01,Kus08}; 
however, we halve 
the lattice constant $b$ 
because we neglect the anion order. 
Therefore, we assume that 
\mbox{$a = 7.083~{\mathrm \AA}$}, 
\mbox{$b = 7.667~{\mathrm \AA}$}, 
\mbox{$c = 13.182~{\mathrm \AA}$}, 
\mbox{$\alpha = 84.40^{\circ}$}, 
\mbox{$\beta = 87.62^{\circ}$}, and 
\mbox{$\gamma = 69.00^{\circ}$}.

We refer to the lattice vectors along 
the crystal a-, b-, and c-axes 
as $\va$, $\vb$, and $\vc$, respectively. 
We also define 
${\hat \va} = \va/a$,\, 
${\hat \vb} = \vb/b$,\, 
${\hat \vc} = \vc/c$,\, 
$\va^{*} = v^{-1} \vb \times \vc$,\, 
$\vb^{*} = v^{-1} \vc \times \va$,\, 
$\vc^{*} = v^{-1} \va \times \vb$, 
and 
$v = \va \cdot (\vb \times \vc)$. 
The crystal momentum $\vk$ can be expressed 
as $\vk = k_x \va^{*} + k_y \vb^{*} + k_z \vc^{*}$, 
where 
$k_x = \vk \cdot \va$,\, 
$k_y = \vk \cdot \vb$,\, 
and $k_z = \vk \cdot \vc$. 
Similarly, 
the FFLO vector $\vq$ is expressed as 
$\vq = q_x \va^{*} + q_y \vb^{*} + q_z \vc^{*}$. 
We define 
$q_1$ and $\varphi_{\vq}$ by 
$(q_x,q_y) = q_1 \, (\cos \varphi_{\vq}, \sin \varphi_{\vq})$, 
and assume that $q_z = 0$. 
Because $\vb$ is not perpendicular to $\va$, 
$q_1 \ne |\vq| \equiv q$. 
We define the unit vector ${\hat \vb}'$ 
that 
satisfies 
\mbox{${\hat \vb}' \cdot {\hat \va} = 0$} 
and 
\mbox{${\hat \vb}' \cdot \vb > 0$} 
by 
\mbox{${\hat \vb}' = ({\hat \vb}-\cos \gamma \cdot {\hat \va})/\sin \gamma$}. 
Therefore, 
\Equation{eq:phivarphi}
{
     \begin{split}
     q \cos \phi_{\vq} & 
       = \vq \cdot {\hat \va} 
       = \frac{q_1}{a} \cos \varphi_{\vq} , \\[-4pt]
     q \sin \phi_{\vq} & 
       = \vq \cdot {\hat \vb}' 
       = \frac{q_1}{a} 
           \Bigl ( 
               \frac{a}{b} \frac{\sin \varphi_{\vq}}{\sin \gamma}
             - \frac{\cos \varphi_{\vq}}{\tan \gamma}   
           \Bigr ) . 
     \end{split}
     }

\begin{figure}[htbp]
\begin{center}
\begin{tabular}{c}
\includegraphics[width=4.5cm]
{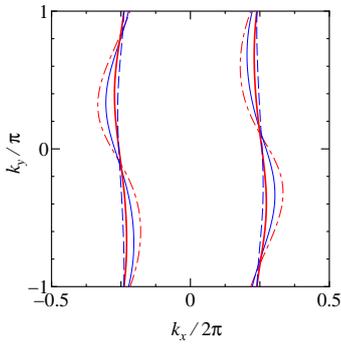}
\\[-4pt]
\end{tabular}
\end{center}
\caption{
(Color online) 
Fermi surfaces derived from \eq.{eq:eps}. 
The red dot-dashed, blue solid, red thick solid, and blue dashed 
curves represent the Fermi surfaces 
for 
parameter sets {\Aave}, {\Pave}, {\Modified1}, and {\Modified2}, 
respectively. 
Although 
$\va^{*}$ and $\vb^{*}$ are not perpendicular to each other, 
the $k_x$-  and $k_y$-axes are drawn perpendicular to each other 
in this figure 
for convenience. 
}
\label{fig:FS}
\end{figure}

Considering application to {\TMTSFClO}, 
we assume the following energy dispersion~\cite{Kis16,Ale14,Pev01}: 
\Equation{eq:eps}
{
     \begin{split}
     \epsilon(\vk) 
     = & 
       - 2 t_{\rm I3} \cos k_y 
       \\[-8pt]
       & 
       - 2 t_{\rm I4} \cos( k_x - k_y ) 
       - \sqrt{ \epsilon_0^2 + [\epsilon_1(\vk)]^2 } , 
     \end{split}
     }
where 
$\epsilon_0^2 = t_{\rm S1}^2 + t_{\rm S2}^2 + t_{\rm I1}^2 + t_{\rm I2}^2$ 
and 
\Equation{eq:eps1}
{
     \begin{split}
     [\epsilon_1(\vk)]^2 
     = & ~ 
         2 t_{\rm S1} t_{\rm S2} \cos k_x 
           + 2 (t_{\rm S1} t_{\rm I1} + t_{\rm S2} t_{\rm I2} ) 
                    \cos k_y 
         \\ 
       & 
           + 2 (t_{\rm S1} t_{\rm I2} + t_{\rm S2} t_{\rm I1} ) 
                    \cos (k_x - k_y) 
       \\ 
       & 
           + 2 t_{\rm I1} t_{\rm I2} \cos ( k_x - 2 k_y) . 
     \end{split}
     }
Physical interpretations of the hopping integrals in real space 
are presented in 
Refs.~\citen{Pev01} and \citen{Kis16}. 
We define 
$\xi(\vk) = \epsilon(\vk) - \mu$, 
$h = \mu_{\rm e} |\vH|$, 
and 
$\xi_{\sigma}(\vk,h) = \xi(\vk) - \sigma h$, 
where 
$\mu$ and $\mu_{\rm e}$ are 
the chemical potential and magnetic moment of the electron, 
respectively. 
We assume a half-filled hole band, 
which corresponds to a quarter-filled hole band 
in a system where the TMTSF molecules are not dimerized. 
We employ the parameter sets 
shown in Table~\ref{table:hopping}, 
where averages are taken because the anion order is neglected. 
The Fermi surfaces for those parameter sets 
are similar, in the sense that they warp in the same directions, 
although the warping magnitudes are different, 
as depicted in Fig.~\ref{fig:FS}.

\begin{table}[hbtp]
\caption{
Assumed parameter sets for {\TMTSFClO} in units of meV. 
Parameter sets~{\PA} and {\PB} 
are 
the sets of hopping integrals obtained 
by {\Pevelen} et al.~\cite{Pev01} 
from their X-ray crystallography results 
for the molecules that are nonequivalent as a result of the anion order. 
{\Pave} is the parameter set obtained from {\PA} and {\PB} 
by simple averaging, 
where the small differences between $t_{\rm I3}$ and $t_{\rm I5}$ 
and between $t_{\rm I4}$ and $t_{\rm I6}$, 
which are caused by the anion order, 
are ignored. 
{\Aave}
is the parameter set obtained by Alemany et al.~\cite{Ale14} 
Parameter sets {\Modified1} and {\Modified2} are similar to {\Pave}, 
but the inter-chain hopping integrals are modified 
so that the experimental results are precisely reproduced. 
}
\label{table:hopping}
{\footnotesize 
\begin{center}
\begin{tabular}{l}
\\
\begin{tabular}{c rrrrrr}
\hline 
  & \multicolumn{1}{c}{{\PA}} 
  & \multicolumn{1}{c}{{\PB}} 
  & \multicolumn{1}{c}{{\Pave}} 
  & \multicolumn{1}{c}{{\Aave}} 
  & \multicolumn{1}{c}{{\Modified1}} 
  & \multicolumn{1}{c}{{\Modified2}} 
  \\ 
\hline 
$t_{\rm S1}$ & 413  & 362  & 387.5   & 278.5  & 387.5  &  387.5 \\
$t_{\rm S2}$ & 324  & 335  & 329.5   & 242.6  & 329.5  &  329.5 \\
$t_{\rm I1}$ 
             & \multicolumn{2}{c}{$-50$} 
                    & $- 50$  & $-50.1$  & $- 50$  &  $- 50$  \\
$t_{\rm I2}$ 
             & \multicolumn{2}{c}{$-100$} 
                    & $-100$  & $-56.4$  & $-90 $  &  $- 90$  \\
$t_{\rm I3}$ 
(/$t_{\rm I5}$) 
             & \multicolumn{2}{c}{70/71} 
                    & 70.5    & 55.9     & 55   &  55  \\
$t_{\rm I4}$ 
(/$t_{\rm I6}$)
             & \multicolumn{2}{c}{20/21} 
                    & 20.5    & $-2.4$   & 35   &  45  \\
\hline 
\end{tabular}
\end{tabular}
\end{center}
}
\end{table}

{\it Formulation} -- 
The transition temperature and upper critical field 
can be calculated on the basis of equations provided 
in previous papers~\cite{Shi94,Shi97a,Shi99aT0,Shi99aFT,Miy14,NoteEqs}. 
In this letter, 
we calculate $h_{\rm c}(\phi) = \mu_{\rm e} H_{\rm c}(\phi)$ 
at $T = 0$. 
As previously mentioned, 
the direction of $\vq$ is locked in the direction of $\vH$; 
i.e., $\phi_{\vq} = \phi$, 
whereas 
the length $q$ must be optimized so that $h_{\rm c}$ is maximized. 
The $\phi$ dependence of 
$h_{\rm c}(\phi)$ at a fixed $T$ is different from that of 
$T_{\rm c}(\phi)$ at a fixed $|\vH|$, 
and the orbital pair-breaking effect 
must significantly reduce the magnitude of $h_{\rm c}$; 
however, 
it is useful to examine $h_{\rm c}(\phi)$ 
because it represents the extent of the stability of the FFLO state 
caused by the nesting effect.

In Q1D systems with open Fermi surfaces, 
we define $s = {\rm sgn}(k_x)$. 
The gap function near the Fermi surfaces 
(\mbox{$k_x = s k_{{\rm F}x}(k_y)$}) 
is expressed as 
$\Delta(s,k_y) = \Delta_{\alpha} \gamma_{\alpha}(s,k_y)$, 
where $\alpha$ is the symmetry index. 
We examine both s-wave and d-wave states 
expressed by $\gamma_{\rm s}(k_y) = 1$ 
and 
$\gamma_{\rm d}(k_y) = \dps{ \sqrt{2} \cos k_y }$, 
respectively, 
although 
in {\TMTSFClO}, presumably, 
the d-wave is the most likely pairing symmetry.~\cite{Notecosky} 
We do not consider 
the possibility of triplet states 
in this letter.~\cite{NoteFFLOTMTSF,Shi00tri,Shi00mix,Aiz09}

\begin{figure}[htbp]
\begin{center}
\begin{tabular}{c}
\includegraphics[width=6.0cm]
{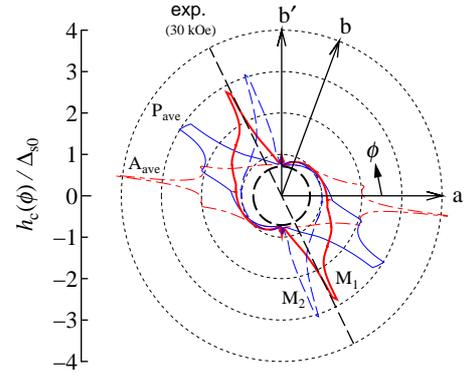}
\\[-4pt]
\end{tabular}
\end{center}
\caption{
(Color online) 
Upper critical fields of s-wave FFLO state at \mbox{$T = 0$} 
scaled by zero-field gap $\Delta_{{\rm s}0}$. 
The red dot-dashed, blue solid, red thick solid, and blue dashed 
curves show 
the results 
for 
parameter sets {\Aave}, {\Pave}, {\Modified1}, and {\Modified2}, 
respectively. 
The dashed straight line represents the direction of the magnetic field 
($- 63.3^{\circ}$) 
at which the experimental $T_{\rm c}^{\rm onset}(\phi)$ has a maximum value 
for $H = 30~{\rm kOe}$.~\cite{Yon08a} 
The bold dashed circle shows the Pauli paramagnetic limit 
$h_{\rm P}/\Delta_{\rm s0} = 1/\sqrt{2} \approx 0.7071$. 
}
\label{fig:hc_swave}
\end{figure}

\begin{figure}[htbp]
\begin{center}
\begin{tabular}{c}
\includegraphics[width=6.0cm]
{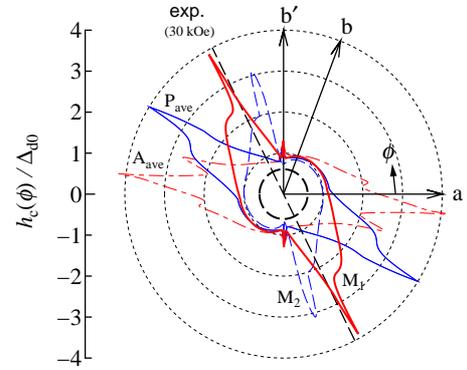}
\\[-4pt]
\end{tabular}
\end{center}
\caption{
(Color online) 
Upper critical fields of d-wave FFLO state at $T = 0$. 
The legend of this figure is 
similar to that of Fig.~\ref{fig:hc_swave}, 
except for the pairing symmetry 
and the values of the Pauli paramagnetic limit 
$h_{\rm P}/\Delta_{\rm d0}$, 
which are approximately equal to 
$0.6075$, $0.6096$, $0.6063$, and $0.6070$ 
for {\Pave}, {\Aave}, {\Modified1}, and {\Modified2}, 
respectively, 
whose curves coincide within the line width. 
}
\label{fig:hc_dwave}
\end{figure}

{\it Results} -- 
Figures~\ref{fig:hc_swave} and~\ref{fig:hc_dwave} show 
the $h_{\rm c}(\phi)$ of the s-wave and d-wave FFLO states, respectively. 
Over wide ranges of $\phi$, 
the upper critical fields $h_{\rm c}(\phi)$ 
are remarkably enhanced by the emergence of the FFLO state. 
In particular, they exhibit sharp cusps, 
at the tops of which $h_{\rm c}$ is more than 
six times the Pauli paramagnetic limit.~\cite{Notecusp} 
Their sharpness implies that the directions of the cusps 
must remain the optimum directions of the magnetic field 
that stabilize the FFLO state the most 
when the orbital pair-breaking effect is incorporated. 
The optimum directions are sensitive to 
changes in the inter-chain hopping integrals, 
whereas for all parameter sets, 
they are in the second and fourth quadrants, 
which do not contain the directions of $\pm \vb$. 
This agrees with the observations in $T_{\rm c}^{\rm onset}(\phi)$. 
The parameter sets {\Modified1} and {\Modified2} 
give the maxima of $h_{\rm c}(\phi)$ 
near 
$\phi \approx 63.3^{\circ}$ and 
$78.3^{\circ}$, 
at which 
the experimental $T_{\rm c}^{\rm onset}(\phi)$ 
have the maximum values when $H = 30$ and $47.5~{\rm kOe}$, respectively. 
Comparing Figs.~\ref{fig:hc_swave} and~\ref{fig:hc_dwave}, 
it is found that 
the optimum directions of $\phi$ 
do not strongly depend on the pairing symmetry.

At the cusps, it is easily verified that \mbox{$n = 3$}, 
which means the terms in $\Delta k_{{\rm F}x}(k_y,\vq)$ 
proportional to $(k_y - k_y^0)^1$ and $(k_y - k_y^0)^2$ vanish. 
This behavior is essentially the same as 
that in the square lattice system mentioned above,~\cite{Shi99aT0} 
although the controlling parameters ($\phi$ and $n_{\rm h}$) 
are different.

\begin{figure}[htbp]
\begin{center}
\begin{tabular}{c}
\includegraphics[width=6.0cm]
{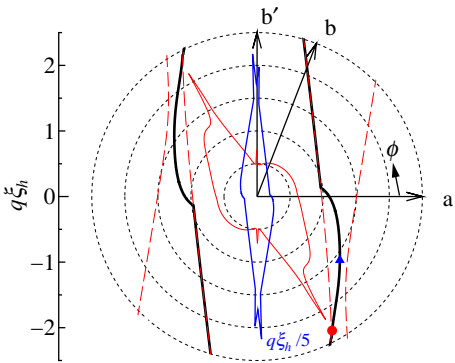}
\\[-4pt]
\end{tabular}
\end{center}
\caption{
(Color online) 
Angular dependence of optimum $q(\phi)$ 
for parameter set {\Modified1} and d-wave pairing. 
$\xi_h(\phi) \equiv v_{\rm F}^{0}/2h_{\rm c}(\phi)$ 
and 
$v_{\rm F}^{0} \equiv 
  t_{\rm S1} t_{\rm S2} a
  /\hbar [t_{\rm S1}^2 + t_{\rm S2}^2]^{1/2}$ 
are defined.~\cite{NotevF0}
The blue solid curve shows the value divided by five. 
The red dashed curves show the $q(\phi)$ values 
that make the Fermi surfaces touch. 
The red closed circle and blue closed triangle 
show the optimum $q(\phi)$ 
at $\phi = -0.340 \pi$ and $-0.210 \pi$, 
respectively. 
For these points, the Fermi surfaces are depicted 
in Figs.~\ref{fig:FSnesting}(a) and (b) 
and Fig.~\ref{fig:FSnesting}(c), 
respectively. 
The former is close to the $\phi$ at which $h_{\rm c}(\phi)$ is a maximum. 
For reference, 
the critical field is shown by the red thin solid curve 
in an arbitrary scale. 
}
\label{fig:FFLOvector}
\end{figure}

\begin{figure}[htbp]
\begin{center}
\begin{tabular}{ccc}
& \\
\includegraphics[height=4.5cm]
{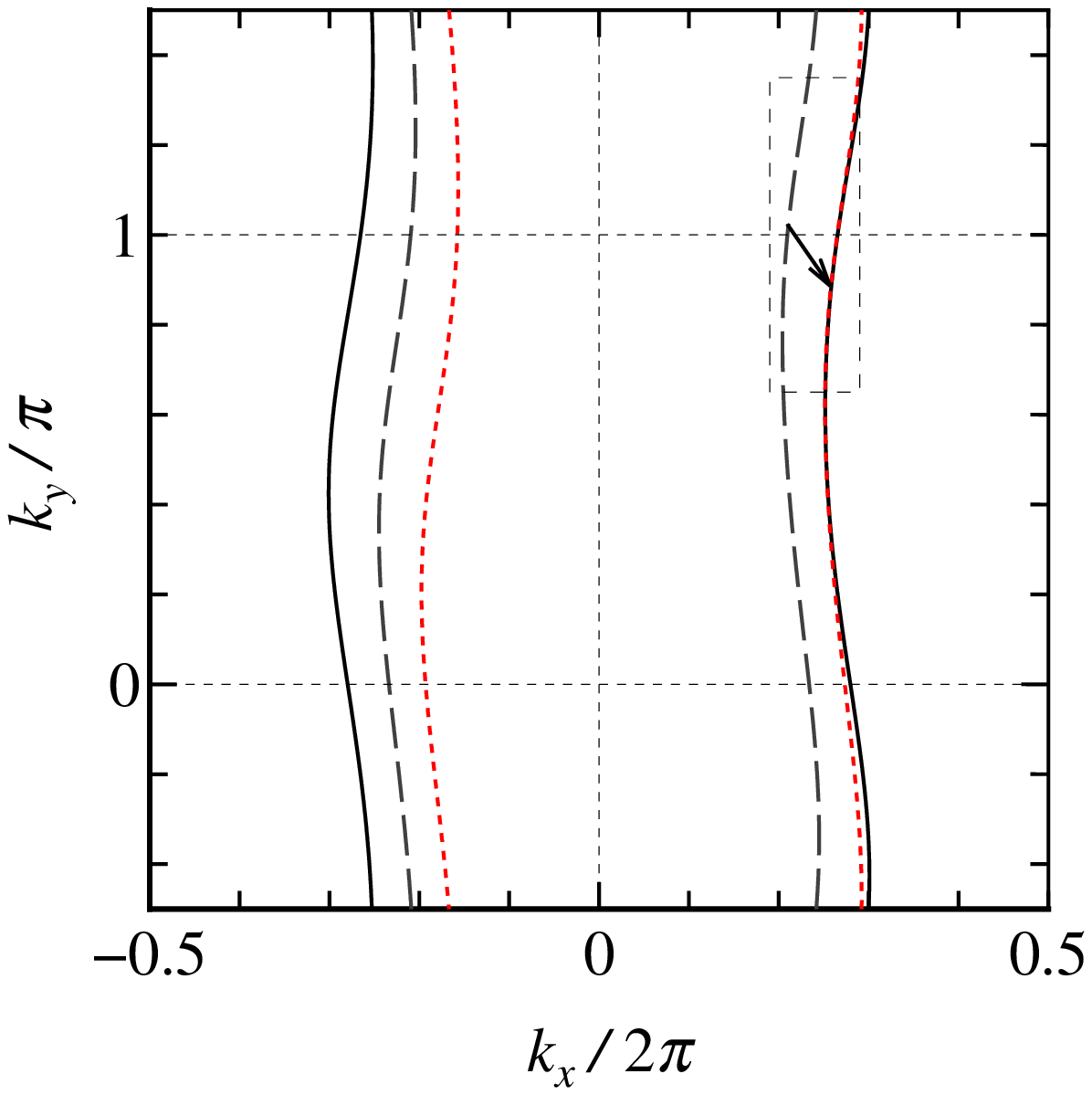}
& 
\includegraphics[height=4.5cm]
{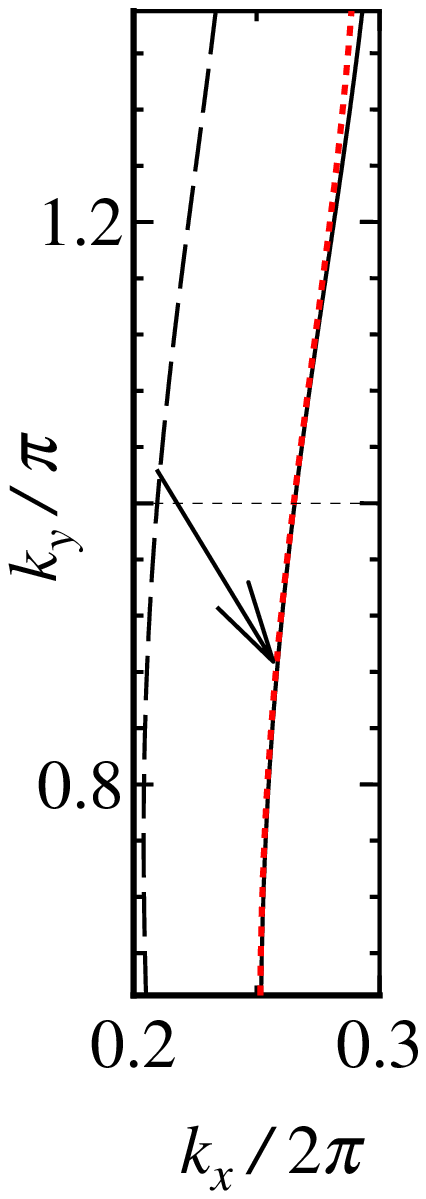}
& 
\includegraphics[height=4.5cm]
{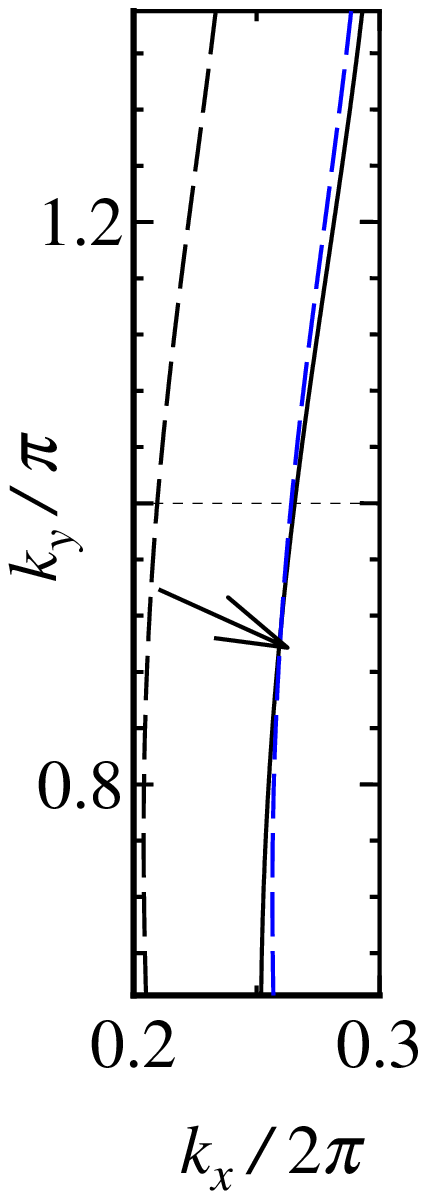}
\\
{\footnotesize (a)}
& 
{\footnotesize (b)}
& 
{\footnotesize (c)}
\\[-8pt]
\end{tabular}
\end{center}
\caption{
(Color online) 
Fermi-surface nesting 
when parameter set {\Modified1} is assumed. 
In panels (a) and (b) and panel (c), 
$\phi_{\vq} = - 0.340 \pi$ and $- 0.210 \pi$ 
(i.e., 
$\varphi_{\vq} = - 0.3070 \pi$ and $-0.1200 \pi$), 
for which 
$h_{\rm c} \approx 3.895 \Delta_{\rm d0}$ and 
$1.738 \Delta_{\rm d0}$, respectively. 
The solid and dashed curves represent 
the Fermi surfaces of the holes with up and down spins, 
respectively. 
The red thin dotted and blue thin dashed curves 
represent the Fermi surfaces of the holes 
with down spins that are shifted by the vectors 
$\vq$ (the small arrows), 
whose lengths maximize $h_{\rm c}$ 
for the given values of $\phi$. 
The $k_x$- and $k_y$-axes are drawn perpendicular to each other. 
In the present weak coupling theory, $h \ll t_{\rm S1}$; 
however, 
for this figure, 
we used a large value $h = 0.1 t_{\rm S1}$ 
to make the displacement visible. 
}
\label{fig:FSnesting}
\end{figure}

In Fig.~\ref{fig:FFLOvector}, 
it is found that at each $\phi$, 
the Fermi surfaces touch or nearly touch 
when $q$ is optimized. 
At $\phi \approx - 0.340 \pi$, 
which is close to the optimum $\phi$, 
the red closed circle shows that 
the optimum $q$ makes the Fermi surfaces touch. 
For this point, 
Figs.~\ref{fig:FSnesting}(a) and (b) 
depict the Fermi surfaces and optimum $\vq$ 
(the given $\phi$ and optimum $q$). 
Interestingly, 
at a $\phi$ value for which the two red thin dashed curves are very close, 
such as at $\phi \approx -0.210 \pi$ (the blue closed triangle), 
the optimum $q$ deviates from those curves, 
which means the Fermi surfaces cross 
for the optimum $q$. 
However, 
as shown in Fig.~\ref{fig:FSnesting}(c), 
those crossing Fermi surfaces are very close 
over a wide range of $k_y$, 
and their crossing angle is extremely small. 
Therefore, for most practical purposes, 
we can regard the Fermi surfaces as touching when $q$ is optimized.

Figure~\ref{fig:FSnesting} also shows that 
the points at which the Fermi surfaces touch 
are far away from $k_y = \pi/2$, 
near which the anion order affects the electron dispersion. 
Hence, the anion order would not significantly 
change the present result.~\cite{Miy16}.

{\it Discussion} -- 
The discrepancy between 
the theoretical and experimental results 
is due to the simplifications in the present theory 
and lack of accurate information on the model parameters. 
For example, 
although the optimum direction of $\vH$ depends on 
the magnitude of the magnetic field 
in the experimental observations, 
that is not so in the present theory. 
This discrepancy may be improved 
if the orbital pair-breaking effect and order-parameter mixing~\cite{Shi00mix} 
are incorporated. 
A more precise analysis that incorporates these factors 
is left for future research.

Although we found parameter sets consistent 
with the observations, 
the ranges of the parameters that 
reproduce the experimental results 
have not been clarified. 
The relation between the optimum direction and hopping parameters 
will be examined in a separate paper.

{\it Conclusion} -- 
In Q1D systems, 
the FFLO state is extremely stable 
for in-plane field directions around the nontrivial optimum directions 
indicated by the cusps in $h_{\rm c}(\phi)$. 
Interestingly, 
this behavior with cusps where $\phi$ is controlled 
is analogous to that in a square lattice system 
in which $n_{\rm h}$ is controlled to deform the Fermi surfaces. 
Hence, 
a similar behavior can occur in other low-dimensional systems 
with other controlling parameters. 
For {\TMTSFClO}, 
it was shown that there exist 
realistic parameter sets ({\Modified1} and {\Modified2}) 
that can reproduce 
the optimum directions of $\vH$ ($\parallel \vq$) 
consistent with the experimental observations. 
Furthermore, 
for the parameter sets obtained from previous studies ({\Pave} and {\Aave}), 
the optimum directions are in the quadrants consistent 
with the experimental observations. 
These results are consistent with 
the hypothesis that 
the FFLO state emerges in the Q1D organic superconductor {\TMTSFClO}.

\mbox{}


\begin{acknowledgments}

The authors would like to thank 
\mbox{S.\,Yonezawa} for the useful discussions and information. 
The authors would also like to thank 
\mbox{K.\,Kishigi} for the useful discussions.

\end{acknowledgments}




\begin{thebibliography}{99}



\bibitem{Ful64}
  P. Fulde and R. A. Ferrell, Phys. Rev. {\bf 135}, A550 (1964). 

\bibitem{Lar64}
  A.\,I.\,Larkin and Yu.\,N.\,Ovchinnikov, 
  Zh. Eksp. Teor. Fiz. {\bf 47}, 1136 (1964);
  translation: Sov. Phys. JETP, {\bf 20}, 762 (1965). 

\bibitem{Cas04}
  R. Casalbuoni and G. Nardulli, 
    Rev. Mod. Phys. {\bf 76}, 263 (2004). 

\bibitem{Mat07} 
  Y. Matsuda and H. Shimahara, 
    J. Phys. Soc. Jpn. {\bf 76}, 051005 (2007). 

\bibitem{Leb08} 
  H. Shimahara, in 
  {\it The Physics of Organic Superconductors and Conductors}, 
     ed. A.G. Lebed (Springer, Berlin, 2008), p.~687. 


\bibitem{NoteTMTSF}
  TMTSF stands for tetramethyltetraselenafulvalene. 


\bibitem{Yon08a} 
  S. Yonezawa, S. Kusaba, Y. Maeno, P. Auban-Senzier, C. Pasquier, 
  \mbox{K.\,Bechgaard}, and D. Jerome, 
    Phys. Rev. Lett. {\bf 100}, 117002 (2008). 

\bibitem{Yon08b} 
  S.\,Yonezawa, S.\,Kusaba, Y.\,Maeno, P.\,Auban-Senzier, C.\,Pasquier, 
  and D.\,Jerome, 
    J. Phys. Soc. Jpn. {\bf 77}, 054712 (2008). 


\bibitem{Gru66}
  L. W. Gruenberg and L. Gunther, 
    Phys. Rev. Lett. {\bf 16}, 996 (1966). 



\bibitem{NoteHLL} 
  When the orbital effect is vanishingly weak, 
  the Abrikosov function can have large Landau level indices $n_{\rm L}$. 
  The order parameters with $n_{\rm L} \gg 1$ 
  exhibit a spatial modulation perpendicular to 
  the vortex line. 
  This modulation is physically equivalent to 
  the FFLO modulation 
  because in the limit $n_{\rm L} \rightarrow \infty$, 
  the vortex state is reduced to the FFLO state.~\cite{Shi97b,Shi09} 
  Unless such states with large $n_{\rm L}$ are considered, 
  FFLO modulation 
  can occur only in the direction parallel to $\vH$.~\cite{Gru66} 


\bibitem{Lee95} 
  I.\,J.\,Lee, A.\,P.\,Hope, M.\,J.\,Leone, and M.\,J.\,Naughton, 
    Synthetic Metals {\bf 70}, 747 (1995). 



\bibitem{Yon12} 
  S.\,Yonezawa, Y.\,Maeno, K.\,Bechgaard, and D.\,\Jerome, 
    Phys. Rev. B {\bf 85}, 140502(R) (2012). 


\bibitem{NoteOrb}
  Near $H = 0$, $H_{\rm c2} \propto T_{\rm c}^{(0)} - T$, 
  which indicates the presence of the orbital pair-breaking effect. 



\bibitem{Shi97b}
  H. Shimahara and D. Rainer, 
    J. Phys. Soc. Jpn. {\bf 66}, 3591 (1997). 


\bibitem{Shi99b}  
  H. Shimahara, 
    Journal of Superconductivity, {\bf 12}, 469 (1999). 


\bibitem{Shi09} 
  H. Shimahara, Phys. Rev. B {\bf 80}, 214512 (2009). 



\bibitem{Shi94}  
  H. Shimahara, Phys. Rev. B {\bf 50}, 12760 (1994). 


\bibitem{Shi97a}  
  H. Shimahara, 
    J. Phys. Soc. Jpn. {\bf 66}, 541 (1997). 


\bibitem{Shi99aT0}  
  H. Shimahara, 
    J. Phys. Soc. Jpn. {\bf 68}, 3069 (1999). 


\bibitem{Shi99aFT}  
  H. Shimahara and K. Moriwake,     
    J. Phys. Soc. Jpn. {\bf 71}, 1234 (2002); 
  \mbox{H. Shimahara} and S. Hata,         
    Phys. Rev. B {\bf 62}, 14541 (2000). 


\bibitem{NoteN1D} 
  The term ``nearly'' means that 
  the system has three-dimensional interactions 
  that stabilize the long-range order 
  at finite temperatures. 


\bibitem{NoteFFLOTMTSF} 
  See references in Ref.~\citen{Leb08}. 


\bibitem{Shi00mix}  
  H. Shimahara, 
    Phys. Rev. B {\bf 62}, 3524 (2000). 



\bibitem{Leb11} 
  A.G. Lebed, 
    Phys. Rev. Lett. {\bf 107}, 087004 (2011). 


\bibitem{Leb10} 
  A.G. Lebed and S. Wu, 
    Phys. Rev. B {\bf 82}, 172504 (2010). 


\bibitem{Cro12}  
  M.D. Croitoru, M. Houzet, and A.I. Buzdin, 
    Phys. Rev. Lett. {\bf 108}, 207005 (2012). 

\bibitem{Cro17}  
  For a review, see 
  [M.D.\,Croitoru and A.I.\,Buzdin, 
    Condens. Matter {\bf 2}, 30 (2017)]. 

\bibitem{Miy14}  
  N. Miyawaki and H. Shimahara, 
    J. Phys. Soc. Jpn. {\bf 83}, 024703 (2014). 


\bibitem{Miy16}  
  N. Miyawaki and H. Shimahara, 
    J. Phys.: Conf. Ser. {\bf 702}, 012002 (2016). 



\bibitem{Aiz09} 
  H. Aizawa, K. Kuroki, T. Yokoyama, and Y. Tanaka, 
    Phys. Rev. Lett. {\bf 102}, 016403 (2009). 


\bibitem{Shi00ao}  
  H. Shimahara, 
    Phys. Rev. B {\bf 61}, R14936 (2000). 


\bibitem{NoteMiy16}  
  Also in a simplified model, $\vq \parallel \va$, 
  and the touching of the Fermi surfaces is far away from 
  the zone edge.~\cite{Miy16} 


\bibitem{Pev01}  
  D.\,Le\,{\Pevelen}, J.\,Gaultier, Y.\,Barrans, D.\,Chasseau, 
  F.\,Castet, and L.\,Ducasse, 
    Eur. Phys. J. B {\bf 19}, 363 (2001). 


\bibitem{Kus08}  
  S.\,Kusaba, S.\,Yonezawa, Y.\,Maeno, P.\,Auban-Senzier, 
  C.\,Pasquier, K.\,Bechgaard, and D.\,\Jerome, 
    Solid State Sciences {\bf 10}, 1768 (2008). 



\bibitem{Kis16}
  K. Kishigi and Y. Hasegawa, 
    Phys. Rev. B {\bf 94}, 085405 (2016). 



\bibitem{Ale14}
  P. Alemany, J.-P. Pouget, and E. Canadell, 
    Phys. Rev. B {\bf 89}, 155124 (2014). 


\bibitem{NoteEqs} 
  For the reader's convenience, 
  we present the equations 
  for the second-order transition point $(h,T)$: 
  \begin{equation*}
     \begin{split}
     \log \frac{\Tc^{(0)}}{T} 
       \, = \, & 
       \int_0^{\infty} \hsp{-0.5} d t \,\, 
       \sum_{s = \pm}
       \int_{-\pi}^{\pi} 
       \frac{d k_y}{2 \pi} \frac{\rho_{\alpha}(0,s,k_y)}{N_{\alpha}(0)} 
     \\
     & \, \, 
     \times \sinh^2 \frac{\beta \zeta}{2} 
       \frac{\tanh t}{t \, ( \cosh^2 t + \sinh^2 (\beta \zeta/2) )}  , 
     \end{split}
  \end{equation*}
  where 
  $\Tc^{(0)}$ is the zero-field transition temperature, 
  and we define 
  \begin{equation*}
  \begin{split}
  \zeta & = 
     \frac{1}{2} {\vector v}_{\rm F} \cdot \vq  - h , \\
  \rho_{\alpha}(0, s, k_y) 
        & = 
         \rho (0, s, k_y) 
           \bigl [ \gamma_{\alpha} (s, k_y) \bigr ]^2 , \\
  N_{\alpha}(0) 
        & = 
         \sum_{s = \pm} 
         \int_{-\pi}^{\pi} \rho_{\alpha}(0,s,k_y) \frac{d k_y}{2 \pi} . 
  \end{split}
  \end{equation*}
  $\rho(\xi,s,k_y)$ is the density of states defined by 
  \begin{equation*}
       \frac{1}{N} \sum_{\vk} F(\vk)
         = \int d \xi 
           \sum_{s = \pm} 
           \int_{-\pi}^{\pi} \frac{d k_y}{2 \pi} 
             \rho(\xi,s,k_y) F(\xi,s,k_y)
  \end{equation*}
  for the arbitrary smooth function 
  $F(\xi(\vk),s,k_y) = F(\vk)$. 
  In the limit of $T = 0$, 
  $h_{\rm c}$ is the solution of 
  \begin{equation*}
     \begin{split}
       h_{\rm c} 
         & = \frac{1}{2} \Delta_{\alpha 0} \exp( f_{\alpha}(\vq) ) , \\ 
       f_{\alpha}(\vq) 
         & = 
             - \sum_{s = \pm}
               \int_{-\pi}^{\pi} 
               \frac{d k_y}{2 \pi} 
               \frac{\rho_{\alpha}(0,s,k_y)}{N_{\alpha}(0)} 
             \log \left | 
                1 - \frac{{\vector v}_{\rm F} \cdot \vq}{2 h_{\rm c}} 
                  \right | , 
     \end{split}
  \end{equation*}
  where 
  $\Delta_{\alpha 0}$ denotes 
  $\Delta_{\alpha}$ 
  when $T = 0$, $H=0$, and $\vq = 0$. 



\bibitem{Notecosky}
  The momentum dependence 
  $\gamma_{\rm d}(k_y) 
  \equiv \gamma_{\rm d}(k_{{\rm F}x}(k_y),k_y) = \sqrt{2} \cos k_y$ 
  may appear to originate from inter-chain pairing. 
  However, in reality, it comes from intra-chain pairing. 
  As derived in Ref.~\citen{Shi89}, 
  the d-wave state induced by the spin fluctuations 
  in the quarter-filled band 
  is primarily expressed 
  as $\gamma_{\rm d}(k_x,k_y) \propto \cos 2 k_x$ 
  (i.e., 
  $\gamma_{\rm d}(k_x,k_y) \propto \cos k_x$ 
  when the molecules are dimerized). 
  The momentum dependence 
  $\gamma_{\rm d}(k_y) \propto \cos k_y$ 
  simulates the structure of the gap function of the d-wave state 
  near the Fermi surface. 
  We confirmed by numerical calculations 
  that this detail does not significantly affect the result. 



\bibitem{Shi00tri}  
  H. Shimahara, 
    J. Phys. Soc. Jpn. {\bf 69}, 1966 (2000). 


\bibitem{Notecusp} 
  For example, for parameter set {\Modified1}, 
  the cusps occur at 
  $\phi \approx -0.3392 \pi$ and $0.6608 \pi$ 
  (i.e., $\varphi_{\vq} \approx - 0.3067 \pi$ and $0.6933 \pi$). 
  The maximum value is 
  $h_{\rm c} \approx 3.925 \Delta_{\rm d0}$, which is given by 
  $q \approx 5.216 \times h/t_{\rm S1}$. 


\bibitem{NotevF0} 
  $v_{\rm F}^{0}$ is the Fermi velocity 
  at the half-filling in the 1D system 
  with $t_{\rm I1} = t_{\rm I2} = t_{\rm I3} = t_{\rm I4} = 0$. 


\bibitem{Shi89} 
   H. Shimahara, 
    J. Phys. Soc. Jpn. {\bf 58}, 1735 (1989). 




\end{thebibliography}
\end{document}